\documentclass[a4paper,parskip=half,captions=tableabove,11pt,toc=bibliography]{scrartcl}
\pdfoutput=1

\usepackage[utf8]{inputenc}
\usepackage{lmodern}
\usepackage[british]{babel}
\usepackage{csquotes}
\usepackage{amsmath,amssymb,mathtools}
\usepackage[separate-uncertainty=true,retain-unity-mantissa=false]{siunitx}
\usepackage{braket}
\usepackage{esdiff}
\usepackage{graphicx}
\usepackage{booktabs,multicol}
\usepackage{listings}
\usepackage[breaklinks, hidelinks]{hyperref}
\usepackage{breakurl}
\usepackage{cleveref}
\usepackage[nohyperlinks]{acronym}
\usepackage{xspace}
\usepackage{authblk}
\usepackage{todonotes}
\usepackage[sorting=none,%
  citestyle=numeric-comp,%
  bibstyle=mynumeric,%
  giveninits=true]{biblatex}

\graphicspath{{graphics/}}
\NewBibliographyString{refname}
\NewBibliographyString{refsname}
\DefineBibliographyStrings{english}{%
  refname = {ref\adddot},
  refsname = {refs\adddot}
}
\DeclareCiteCommand{\ccite}
  {%
  \ifnum\thecitetotal=1
    \bibstring{refname}%
  \else%
    \bibstring{refsname}%
  \fi%
  \addnbspace\bibopenbracket%
  \usebibmacro{cite:init}%
   \usebibmacro{prenote}}
  {\usebibmacro{citeindex}%
   \usebibmacro{cite:comp}}
  {}
  {\usebibmacro{cite:dump}%
   \usebibmacro{postnote}%
   \bibclosebracket}
\newrobustcmd*{\Ccite}{\bibsentence\ccite}

\DeclareSIUnit{\pb}{pb}
\DeclareSIUnit{\fb}{fb}

\acrodef{SM}{Standard Model}
\acrodef{BSM}{beyond the SM}
\acrodef{DM}{dark matter}
\acrodef{EDM}{electric dipole moment}
\acrodef{2HDM}{two-Higgs-doublet model}
\acrodef{R2HDM}{real 2HDM}
\acrodef{C2HDM}{complex 2HDM}
\acrodef{N2HDM}{next-to 2HDM}
\acrodef{TRSM}{two-real-singlet-extension of the SM}
\acrodef{CxSM}{complex-singlet-extension of the SM}
\acrodef{CL}{confidence level}
\acrodef{EW}{electroweak}
\acrodef{h125}[$h_{125}$]{the \SI{125}{GeV} Higgs boson}
\acrodef{vev}{vacuum expectation value}
\acrodef{EWSB}{EW symmetry breaking}
\acrodef{EWPT}{EW phase transition}

\newcommand{\eqdot}{\,.}
\newcommand{\eqcomma}{\,,}
\newcommand{\hc}{\text{h.c.}}

\newcommand{\ie}{\textit{i.e.}~}
\newcommand{\eg}{\textit{e.g.}~}

\newcommand{\CL}[1]{\ensuremath{\SI{#1}{\%}~\text{\acs{CL}}}\xspace}

\newcommand{\Scanners}{\textsf{ScannerS}\xspace}
\newcommand{\Scannersv}[1]{\mbox{\textsf{ScannerS--#1}}\xspace}
\newcommand{\Evade}{\textsf{EVADE}\xspace}
\newcommand{\Hb}{\textsf{HiggsBounds}\xspace}
\newcommand{\Hbcite}{\Hb~\cite{Bechtle:2008jh,Bechtle:2011sb,Bechtle:2013wla,Bechtle:2015pma,Bechtle:2020pkv}\xspace}
\newcommand{\Hs}{\textsf{HiggsSignals}\xspace}
\newcommand{\Hscite}{\Hs~\cite{Bechtle:2013xfa,HS2Manual}\xspace}
\newcommand{\anyHdecay}{\textsf{anyHdecay}\xspace}
\newcommand{\Bsmpt}{\textsf{BSMPT}\xspace}
\newcommand{\Mo}{\textsf{MicrOMEGAs}\xspace}
\newcommand{\Mocite}{\Mo~\cite{Belanger:2006is,Belanger:2008sj,Belanger:2010gh,Belanger:2013oya,Belanger:2014vza,Barducci:2016pcb,Belanger:2018mqt}\xspace}
\newcommand{\SUL}{\ensuremath{{\mathrm{SU}(2)}_L}\xspace}
\newcommand{\Ztwo}{\ensuremath{\mathbb{Z}_2}\xspace}
\newcommand{\Mathematica}{\textsf{Mathematica}\xspace}

\title{\Scanners{} --- Parameter Scans\\ in Extended Scalar Sectors}
\author{Margarete Mühlleitner}
\affil{Institute for Theoretical Physics,
       Karlsruhe Institute of Technology,
       76128~Karlsruhe,
       Germany}
\author{Marco O. P. Sampaio}
\affil{Feedzai --- Porto,
       Rua de Santos Pousada 228 Piso 4,
       4000-278 Porto,
        Portugal}
\author[3,4]{Rui Santos}
\affil[3]{Centro de Física Teórica e Computacional,
       Faculdade de Ciêncas, Universidade de Lisboa,
       Campo Grande,
       Edifício C8,
       1749-016~Lisboa,
       Portugal}
\affil[4]{ISEL --- Instituto Superior de Engenharia de Lisboa,
       Instituto Politécnico de Lisboa,
       1959-007~Lisboa,
       Portugal}
\author[5]{Jonas Wittbrodt}
\affil[5]{Department of Astronomy and Theoretical Physics,
       Lund University,
       Sölvegatan~14A, 223~62~Lund,
       Sweden}
\date{}
\titlehead{\hfill \parbox{3cm}{KA-TP-05-2020\\ LU TP 20-38}}

\begin{document}
\maketitle
\thispagestyle{empty}
\setcounter{page}{0}
\begin{abstract}
  We present the public code \Scannersv{2} that performs parameter scans and
  checks parameter points in theories beyond the Standard Model (BSM) with
  extended scalar sectors. \Scanners incorporates theoretical and experimental
  constraints from many different sources in order to judge whether a parameter
  point is allowed or excluded at approximately \CL{95}. The BSM models
  implemented in \Scanners include many popular BSM models such as singlet
  extensions, different versions of the Two-Higgs-Doublet Model, or the
  different phases of the Next-to Two-Higgs-Doublet Model. The \Scanners
  framework allows straightforward extensions by additional constraints and BSM
  models.
\end{abstract}
{\let\thefootnote\relax\footnotetext{Electronic addresses: \href{mailto:milada.muehlleitner@kit.edu}{milada.muehlleitner@kit.edu}, \href{mailto:marco.sampaio@feedzai.com}{marco.sampaio@feedzai.com}, \href{mailto:rasantos@fc.ul.pt}{rasantos@fc.ul.pt}, \href{mailto:jonas.wittbrodt@thep.lu.se}{jonas.wittbrdt@thep.lu.se}}}

\clearpage
\tableofcontents

\section{Introduction}\label{sec:intro} In the exploration of extensions of the
\ac{SM} of particle physics it is mandatory to work with \emph{allowed}
parameter points --- points within the parameter space of the model that do not
disagree with current observations. Such samples of allowed parameter points can
be used to showcase the possible phenomenology in a \ac{BSM} theory or to
illustrate the impact of novel calculations for a physically relevant scenario.
The code \Scanners~\cite{Coimbra:2013qq} can be used to generate and validate
samples of such allowed parameter points in many \ac{BSM} models with extended
scalar sectors. \Scanners performs \emph{parameter scans} within the parameter
space of the model.  A \emph{parameter scan} is a greatly simplified \emph{model
  fit} that forgoes making any statements about the model as a whole in favor of a
simpler, point-by-point approach.

In \emph{model fits} --- such as those you can perform with the
\textsf{Fittino}~\cite{Bechtle:2004pc}, \textsf{ZFitter}~\cite{Arbuzov:2005ma},
\textsf{GFitter}~\cite{Flacher:2008zq}, \textsf{GAMBIT}~\cite{Athron:2017ard},
or \textsf{HEPFIT}~\cite{deBlas:2019okz} public codes --- an overall combined
$\chi^2$ or likelihood is constructed from all available measurements. By
sampling this distribution and using the resulting best fit point of the model,
statements about the favoured regions of parameter space and the compatibility
between model predictions and data can be made.\footnote{The exact statements
  possible depend on the chosen statistical interpretation --- frequentist,
  bayesian, or some mixture of the two.} However, there are many cases where no
\emph{global} statements about the model are required. In these cases performing
a full model fit is often excessive and a simpler approach is sufficient.

The \emph{parameter scan} approach employed in \Scanners does not construct a
global likelihood distribution but instead uses a set of individual
\emph{constraints}. This means, that \Scanners tests the model predictions for
the (randomly generated) input parameter points against all implemented
\emph{constraints} and treats each parameter point that passes all constraints
as \emph{allowed}. This point-by-point approach means that no overall
best-fit-point is found, nor is the resulting sample of allowed parameter points
a faithful representation of the global likelihood distribution. These
limitations mean that the results of a parameter scan should never be used to
make global statements about the model --- such as \textit{``the model fits the
  data well''} or \textit{``the model cannot explain this observation''}.
Furthermore, no conclusions should be drawn based on the density of the allowed
parameter points in the parameter space of the model. This would require the
points to follow a statistically meaningful distribution while, in a parameter
scan, their distribution entirely depends on the sampling.

On the other hand, the results of a \emph{parameter scan} are perfectly suited
for phenomenological benchmark scenarios, illustrating interesting signatures,
checking the effects of precision calculations, and many other applications that
only rely on the existence of allowed parameter points in some region of
parameter space. Since the naive combination of the individual constraints leads
to an overestimated combined constraint, the region covered by the allowed
parameter points is a conservative estimate for the favoured region that would
be found in a model fit.\footnote{This only holds as long as correlations
  between different constraints are small. In \Scanners, most constraints stem
  from distinct sources and can be assumed uncorrelated to a good approximation.}
The most important advantage of \emph{parameter scans} is simplicity. Since no
best fit point is found or needed\footnote{In the constraints that are
  reconstructed as a $\chi^2$ value --- notably the oblique parameters and the
  Higgs measurements --- \Scanners applies the constraint with the \ac{SM} as
  alternative hypothesis.} it is perfectly fine to run small parameter scans that
only yield a few \emph{allowed} parameter points or to focus on some special
regions of the parameter space that may not necessarily contain the best fit
point of the model. Additionally, the non-reliance on the best fit point means
that an existing sample of parameter points can simply be re-checked with
\Scanners if some constraints are updated and no new scan is required.

This manual discusses the physics of \Scanners and accompanies the release of
\Scannersv{2}. It is complemented by the technical documentation available
online at
\begin{center}
  \url{https://jonaswittbrodt.gitlab.io/ScannerS}\eqdot{}
\end{center}
\Scannersv{2} is a new code inspired by the old
\Scannersv{1}~\cite{Coimbra:2013qq}. Compared to the old code, we implemented
substantial technical upgrades and many physics improvements. In this manual we
first discuss the constraints included in \Scanners in \cref{sec:constraints}.
In \cref{sec:models} we then give short overviews of the implemented \ac{BSM}
models to establish the conventions used in the code. \Cref{sec:user} contains
instructions on how to build and use \Scanners and an explanation on how
additional models and constraints can be implemented. We summarize in
\cref{sec:summary}. In \cref{app:anyhdecay} we describe the \anyHdecay interface
library and in \cref{app:pertun} we discuss the \Mathematica package for finding
perturbative unitarity constraints that is distributed with \Scanners.

\section{Constraints}\label{sec:constraints}
Experimental constraints in \Scanners are implemented at a \ac{CL} of
$\SI{95}{\%}$ or --- almost equivalently --- $2\sigma$ while theoretical
constraints are simple exclusions without an associated statistical
interpretation. As such, the \emph{allowed} parameter points returned by
\Scanners are valid from a theoretical point of view and not excluded by
observations at approximately \CL{95}.\footnote{The overall constraint is in
  fact slightly stronger than \CL{95} due to the naive combination of constraints.
  Additionally, all correlations between constraints are neglected.} Constraints
in \Scanners have one of three different \emph{severities} governing how their
result is handled by the code. The strongest and default severity is
\texttt{apply}, meaning that only parameter points fulfilling the constraints
will be kept as allowed points. If the weakest severity, \texttt{skip}, is set
for a constraint, no calculations associated with it will be performed and
parameter points will not be tested against it. For a constraint with the
intermediate severity, \texttt{ignore}, all associated calculations will be
executed and the results saved to the output. However, parameter points will be
treated as allowed whether they fulfill the constraint or not, while an
additional output value will be added to indicate if the specific constraint was
fulfilled. Practical information on setting severities is given in
\cref{sec:user}.

Some of the following constraints are not applicable to all of the models
implemented in \Scanners. Detailed technical information regarding the
implementation of constraints in \Scanners can be found in the online
documentation for the
\href{https://jonaswittbrodt.gitlab.io/ScannerS/namespaceScannerS_1_1Constraints.html}{\texttt{ScannerS::Constraints}}
namespace. The online documentation also lists all output quantities associated
to the constraints.

\subsection{Theoretical Constraints}
Theoretical constraints are self-consistency requirements on a parameter point
of the model. As such, they do not have a statistical interpretation --- they
are either fulfilled or the parameter point is invalid and excluded.

\subsubsection{Perturbative Unitarity}
If a theory violates unitarity it is strongly coupled and cannot be treated in
the perturbative approach employed in \Scanners. Unitarity constraints, are
obtained by requiring the eigenvalues $\mathcal{M}^i_{2\to2}$ of the $2\to2$
scattering matrix $\mathcal{M}_{2\to2}$ to fulfill
\begin{equation}
  \left|\mathcal{M}_{2\to2}^i\right| \leq 8\pi\eqdot\label{eq:uni}
\end{equation}
The tree-level scattering matrix can be easily constructed in most BSM models
with extended Higgs sectors in the high energy limit~\cite{Lee:1977eg,
  Kanemura:1993hm}. It often has a block-diagonal form that can --- at least
partially --- be diagonalized analytically. Whenever possible the corresponding
closed form solutions for the eigenvalues are implemented in \Scanners.

Tree-level perturbative unitarity provides only a first approximation to the true
unitarity constraint on the model parameters. It may be impacted by
loop-corrections~\cite{Marciano:1989ns} or finite-energy
effects~\cite{Goodsell:2018tti,Goodsell:2018fex,Krauss:2018orw}.

\subsubsection{Boundedness from Below}
Boundedness of the scalar potential from below is a prerequisite to the
existence of a stable vacuum. In \Scanners, analytical conditions that ensure
boundedness from below are implemented, whenever possible. In general, such
conditions can be very challenging to obtain (see \eg \ccite{Ivanov:2018jmz} for
a recent overview of different methods and their applications). However, closed
form solutions have been found for many \ac{BSM} models --- including all of the
models implemented in \Scanners, see \cref{sec:models}.

\subsubsection{Vacuum Stability}
Even if the scalar potential is bounded from below, the \ac{EW} vacuum is not
necessarily the global minimum and additional constraints can arise from the
stability of the \ac{EW} vacuum.

In some models or phases of models, stability of the tree-level \ac{EW} vacuum
has been proven, or analytic conditions ensuring stability have been found (see
also \cref{sec:models}). For the remaining cases, \Scanners offers a link to the
\Evade library~\cite{Hollik:2018wrr, Ferreira:2019iqb, EVADEweb} to test the
(meta)stability of the \ac{EW} vacuum numerically at tree-level.

\subsection{Electroweak Precision Constraints}\label{sec:EWprecision}

Precision measurements of \ac{EW} observables are sensitive to \ac{BSM} loop
effects present in many extended scalar sectors. As long as the \ac{BSM} effects
are fully captured through self-energy corrections to the gauge-boson
propagators --- which holds in all models discussed here --- and the new physics
scale is not too small, these effects can be parametrized through the oblique
parameters $S$, $T$, and $U$~\cite{Peskin:1991sw}. \Scanners does not compute
any \ac{EW} precision observables and instead uses a fit result --- currently
from \ccite{Haller:2018nnx} --- for the values of the oblique parameters as
constraint. This result includes a covariance matrix to account for correlations
among $S$, $T$, and $U$. In order to calculate model predictions for the oblique
parameters \Scanners implements the results of \ccite{Grimus:2007if,
  Grimus:2008nb}, which are valid for models with any number of scalar \SUL
doublets and singlets. Given the model predictions and the fit result, we
calculate a $\chi^2$ value and treat the constraint as fulfilled if $\chi^2 <
  \chi^2_\text{crit}(2\sigma)$.

\subsection{Flavour Constraints}\label{sec:flavor}
Since all of the models currently implemented in \Scanners are naturally flavor
conserving~\cite{Buras:2010mh}, the dominant contributions to most flavor
observables originate in charged Higgs exchange. Since the charged Higgs sector
of all implemented models is either absent or identical to the \ac{2HDM}, we
treat flavour constraints similar to the \ac{EW} precision constraints and use
the fit results of \ccite{Haller:2018nnx} providing $2\sigma$ constraints in the
$m_{H^\pm}$--$\tan\beta$ parameter plane of the \ac{2HDM}. These generalize
trivially to models with additional singlets, such as the \ac{N2HDM}.

\subsection{Higgs Searches and Higgs Measurements}\label{sec:higgs}
Searches for additional scalars as well as measurements of \ac{h125} are among
the most important constraints on extended scalar sectors. \Scanners provides an
interface to the tools \Hbcite and \Hscite to incorporate these constraints. The
\Hb input can either be given in the effective coupling approximation or by
providing all branching ratios and hadronic cross sections to the Higgs bosons.
For simple models, such as pure singlet extensions, the effective coupling
approximation is exact and used by \Scanners.

Otherwise, \Scanners uses the \anyHdecay library (see \cref{app:anyhdecay}) as
an interface to the many adaptations of the code
\textsf{HDECAY}~\cite{Djouadi:1997yw, Harlander:2013qxa, Djouadi:2018xqq} for
different \ac{BSM} models to obtain model predictions for Higgs boson branching
ratios and total widths including state-of-the-art QCD corrections and off-shell
effects. To obtain precise cross section predictions, \Scanners includes a
tabulated parametrization of the NNLO QCD gluon fusion and $b\bar{b}$-associated
Higgs production cross sections at hadron colliders obtained with
\textsf{SusHi}-1.6.1~\cite{Harlander:2012pb, Harlander:2016hcx}. Additionally,
\Scanners uses cross section parametrizations included in \Hb (see
\ccite{Bechtle:2020pkv}), such as the $W^\pm$/$Z$-associated cross sections
calculated using \textsf{VH@NNLO}~\cite{Brein:2012ne, Harlander:2018yio}, as
well as the $t$-associated charged Higgs production cross section at NLO
QCD~\cite{Berger:2003sm, Dittmaier:2009np, Flechl:2014wfa, Degrande:2015vpa,
  deFlorian:2016spz, Degrande:2016hyf}.

\Hb uses this input to check the model predictions against exclusion bounds from
Higgs searches at LEP, TEVATRON and the LHC\@. Using the expected limit
information, only the most sensitive search for each scalar in the model is
applied to obtain an approximate combined $2\sigma$ constraint on the model
parameter space.

\Hs uses the same input as \Hb to calculate a $\chi^2$ value that quantifies the
agreement of the model prediction with up-to-date measurements of the \ac{h125}
properties at the LHC\@. In interpreting this $\chi^2$, \Scanners uses a
profiled likelihood ratio test with the \ac{SM} as the alternative hypothesis.
In the Gaussian approximation the test statistic is
\begin{equation}
  \Delta\chi^2 = \chi^2_\text{Model} - \chi^2_\text{SM}\eqcomma
\end{equation}
where both $\chi^2_\text{Model}$ and $\chi^2_\text{SM}$ are obtained from \Hs.
As discussed in \cref{sec:intro}, we use this test statistic --- instead of \eg
constructing a goodness-of-fit test using $\chi^2_\text{Model}/\text{d.o.f.}$
--- since it allows for a much easier statistical interpretation and does not
require knowledge about the best fit point of the model. The resulting
$\Delta\chi^2$ approximately describes the best-fit region of the parameter
space. The upper bound $\chi^2_\text{crit}$ to be imposed on $\Delta\chi^2$
depends on the desired confidence level and on the number $\nu$ of degrees of
freedom. As stated above, we aim to impose $2\sigma$ constraints in the Gaussian
limit and thus $\Delta\chi^2 < \chi^2_\text{crit}(2\sigma,\nu)$. When presenting
results where all but $n$ model parameters and all nuisance parameters are
profiled over, the set of allowed parameter points are those with $\Delta\chi^2
  < \chi^2_\text{crit}(2\sigma,n)$. In \Scanners, we by default use $n=2$, which
is appropriate for presenting results as scatter plots and in benchmark planes.
This leads to the criterion
\begin{equation}
  \Delta\chi^2 < 6.18\eqcomma
\end{equation}
which corresponds to a $2\sigma$ constraint under the assumption of Gaussian
errors.

\subsection{Electric Dipole Moments}
In CP-violating models the stringent limits on fermionic \acp{EDM} have to be
considered. \Scanners includes a constraint to check the model prediction for
the electron \ac{EDM} against the latest limits by the ACME
collaboration~\cite{Andreev:2018ayy}. Constraints from the neutron or nuclear
\acp{EDM} could easily be included in \Scanners if theoretical predictions for
these quantities are available in the model.

\subsection{\acs{DM} Constraints}\label{sec:DM}
A global symmetry of an extended scalar sector that remains unbroken after
\ac{EWSB} may lead to a dark sector with a stable lightest
particle. Such a particle is a \ac{DM} candidate and constraints from \ac{DM}
searches need to be considered. \Scanners includes an interface to the code
\Mocite to calculate \ac{DM} observables. The relic density
$\omega_c^\text{model}$ predicted by the model is required \emph{not to exceed}
the observed dark matter density $\omega_c$~\cite{Aghanim:2018eyx},
\begin{equation}
  \omega^\text{model}_c \leq \omega_c + 2 \Delta\omega_c\eqcomma
\end{equation}
where $\Delta\omega_c$ is the uncertainty of the measurement. By imposing only
an upper bound we allow for additional contributions to \ac{DM} beyond the one
stemming from the model under consideration.

\Scanners also imposes constraints from dark matter searches using direct
detection --- more specifically the \textsf{XENON1T}
results~\cite{Aprile:2018dbl}. The required dark-matter--nucleon scattering
cross sections are also obtained from \Mo. \Scanners currently does not include
constraints from indirect detection or collider searches for dark particles, as
these are not available as simple, model-independent limits and require
additional effort to interpret them correctly. Also note that for the models
implemented in \Scanners{} --- that all feature Higgs-portal, weakly
interacting, scalar \ac{DM} in the $\si{\GeV}$ range --- the direct detection
limits are typically stronger than indirect or collider searches.

\subsection{A First Order \acs{EW} Phase Transition}\label{sec:EWPT}
When studying models that could facilitate \ac{EW} baryogenesis~\cite{Kuzmin:1985mm, Shaposhnikov:1986jp,
  Trodden:1998ym, Morrissey:2012db} or in the context of gravitational wave
signatures~\cite{Kosowsky:1992rz, Kosowsky:1991ua, Kosowsky:1992vn,
  Kamionkowski:1993fg, Apreda:2001us, Kosowsky:2001xp, Dolgov:2002ra,
  Grojean:2006bp} (see also \ccite{Caprini:2015zlo}) it is interesting to look at
parameter regions where the \ac{EWPT} is (strongly) first order. \Scanners
includes an interface to the \textsf{BSMPT}~\cite{Basler:2018cwe, Basler:2020nrq}
library to calculate the strength of the \ac{EWPT} in several of the implemented
models. When using this constraint, \Scanners requires that the \ac{EWPT} is of
first order and that the \ac{EW} vacuum at zero temperature is the global
minimum of the 1-loop effective potential. This second requirement is a
prerequisite for the phase transition search in \textsf{BSMPT}.

This constraint by default only requires the \ac{EWPT} to be first order without
putting any bound on the strength of the phase transition. However, the critical
temperature $T_c$ and critical \ac{vev} $\omega_c$ are stored in the output,
such that more stringent requirements --- \eg $\omega_c>T_c$ for a strong first
order \ac{EWPT} --- can be imposed in post-processing. Note that in general,
there is no physical requirement that the \ac{EWPT} has to be first order.
Therefore, you should only enable this constraint if you wish to study physics
associated with a first order \ac{EWPT}.

\section{\acs{BSM} Models in \Scanners}\label{sec:models}
\Scanners includes implementations for a variety of \ac{BSM} models with
extended scalar sectors. In this section, we will give short overviews of the
implemented models mainly meant to establish the conventions used in the
\Scanners implementation. Additional technical information on all implemented
models can be found in the online documentation for the model classes in the
\href{https://jonaswittbrodt.gitlab.io/ScannerS/namespaceScannerS_1_1Models.html}{\texttt{ScannerS::Models}}
namespace. The name of the model class and \Scanners executable for each model
is indicated in the title of the subsection with the corresponding model or
phase description.

\subsubsection{Input Parameters and Mass-Ordering}\label{sec:ordering} As the
very first step for each parameter point, \Scanners calculates all model
parameters from a chosen set of input parameters.\footnote{In the old
  \Scannersv{1} this calculation was performed numerically and referred to as the
  generation of a local minimum. Now, the analytic relations between input
  parameters and the remaining model parameters --- \eg between masses and mixing
  angles on the one hand and the parameters of the scalar potential on the other
  --- are implemented in closed form, whenever possible.} The input parameters are
chosen to allow for a physically motivated selection of scan ranges. This means
that physical masses, mixing angles or couplings, and \acp{vev} are used rather
than directly using the parameters of the scalar potential.

In most \ac{BSM} models, scalars of identical charge and CP, $h_i$, are
distinguished based on their mass --- typically by imposing some kind of mass
ordering, \eg $m_{h_i}\leq m_{h_j}$ for $i<j$. In contrast to more complex BSM
models --- such as supersymmetry or composite Higgs models --- the $m_{h_i}$ can
be chosen as input parameters in all of the scalar extensions implemented in
\Scanners. This greatly facilitates phenomenological studies, as one of the
Higgs masses can be set to match the observed mass of \ac{h125}. However, since
the other scalars can in general be either lighter or heavier than \ac{h125} it
is inconvenient to use mass ordered states as input. Instead, \Scanners uses a
basis of input states $h_{a,b,c,\ldots}$ for which no ordering is required.

As an example, consider a model with three mixing scalars. The three by three mixing matrix is
parametrized as
\begin{equation}
  R = \begin{pmatrix}
    c_1 c_2                  & s_1 c_2                  & s_2     \\
    -(c_1 s_2 s_3 + s_1 c_3) & c_1 c_3 - s_1 s_2 s_3    & c_2 s_3 \\
    - c_1 s_2 c_3 + s_1 s_3  & -(c_1 s_3 + s_1 s_2 c_3) & c_2 c_3
  \end{pmatrix}\eqcomma\label{eq:Rmix}
\end{equation}
where $c_{1,2,3} = \cos \alpha_{1,2,3}$ and $s_{1,2,3} = \sin \alpha_{1,2,3}$
for the three mixing angles $-\pi/2 \leq \alpha_{1,2,3} < \pi/2$. This
parametrization is sufficiently general for physics purposes, but is not a full
parametrization of $O(3)$, in particular --- since $\cos\alpha_i\geq0$ for the
chosen angular ranges --- it requires
\begin{align}
  \det R & = +1\eqcomma & R_{11} & \geq 0\eqcomma & R_{33} & \geq 0\eqdot\label{eq:Rconds}
\end{align}
The model input is given in the form of three Higgs masses,
\eg
\begin{align}
  m_{h_a} & = \SI{125}{\GeV}\eqcomma & m_{h_b} & =\SI{300}{\GeV}\eqcomma & m_{h_c}=\SI{50}{\GeV}\eqcomma\label{eq:examplemass}
\end{align}
and values for the three input mixing angles $\alpha^\text{in}_{1,2,3}$ that parametrize
$R^\text{in}$ in the basis
\begin{equation}
  \begin{pmatrix}h_a\\h_b\\h_c\end{pmatrix} = R^\text{in} \begin{pmatrix}\phi_1\\\phi_2\\\phi_3\end{pmatrix}\eqcomma
\end{equation}
where $\phi_{1,2,3}$ are the mixing fields of the scalar potential. To convert
this to the desired basis ${(h_1,h_2,h_3)}^\text{T}$ with
$m_{h_1}<m_{h_2}<m_{h_3}$, a set of row transpositions is applied on the mixing
matrix, such that --- for the example mass values of \cref{eq:examplemass} ---
\begin{equation}
  \begin{pmatrix}h_1\\h_2\\h_3\end{pmatrix}
  \equiv\begin{pmatrix}h_c\\h_a\\h_b\end{pmatrix}
  = \begin{pmatrix}
    R^\text{in}_{c1} & R^\text{in}_{c2} & R^\text{in}_{c3} \\
    R^\text{in}_{a1} & R^\text{in}_{a2} & R^\text{in}_{a3} \\
    R^\text{in}_{b1} & R^\text{in}_{b2} & R^\text{in}_{b3}
  \end{pmatrix}
  \begin{pmatrix}\phi_1\\\phi_2\\\phi_3\end{pmatrix}\eqdot
\end{equation}
In general, the resulting mixing matrix no longer fulfills \cref{eq:Rconds} and
is thus no longer parametrized by \cref{eq:Rmix}. The parametrization can be
restored by physically irrelevant field redefinitions of the form $h_i\to -h_i$
that flip the sign of all elements in a row of the mixing matrix, \ie $h_1\to
  -h_1$ if $R_{11}<0$, $h_3\to -h_3$ if $R_{33}<0$, and finally $h_2\to -h_2$ if
$\det R = -1$.\footnote{The corresponding implementation can be found in
  \href{https://jonaswittbrodt.gitlab.io/ScannerS/namespaceScannerS_1_1Utilities.html}{\texttt{ScannerS::Utilities::OrderedMixMat3d}}.}

This procedure works analogously for other dimensionalities or different
parametrizations of $R$. It can also be used if not all of the masses are input
parameters. In this case the remaining masses are first calculated using $R^\text{in}$
and then the reordering is performed.

The input parametrizations in \Scanners are usually agnostic regarding which of
the $H_a$ is identified with \ac{h125}. Any exceptions to this are stated in the
description of the parametrizations below.

\subsection{The Complex-Singlet-Extension of the \acs{SM}}\label{sec:CxSM}
Pure extension of the \ac{SM} by gauge-singlet scalar fields are the simplest
possible extended scalar sectors. The \ac{CxSM}~\cite{Barger:2008jx,
  Coimbra:2013qq, Costa:2015llh} adds a complex singlet field $\mathbb{S}$ with a
softly broken $U(1)$ symmetry to the \ac{SM}. The implementation in \Scanners
follows \ccite{Coimbra:2013qq,Costa:2015llh} where an additional symmetry under
$\mathbb{S}\to\mathbb{S}^*$ is imposed that forces all model parameters to be
real. The resulting scalar potential is
\begin{equation}
  V_\text{CxSM} = \frac{m^2}{2} \Phi^\dagger \Phi
  + \frac{\lambda}{4} {(\Phi^\dagger \Phi)}^2
  + \frac{\delta_2}{2} \Phi^\dagger \Phi |\mathbb{S}|^2
  + \frac{b_2}{2} |\mathbb{S}|^2
  + \frac{d_2}{4} |\mathbb{S}|^4
  + \left(\frac{b_1}{4}\mathbb{S}^2 + a_1\mathbb{S} + \hc\right)\label{eq:cxsmpot}
\end{equation}
with seven real parameters. The \ac{CxSM} allows for different phases, where
different fields acquire \acp{vev}. Implementations for the \emph{broken phase}
and the \emph{dark phase} are included in \Scanners. The following constraints
are included for both phases with very similar implementations:
\begin{itemize}
  \item The perturbative unitarity constraint was obtained purely numerically in
        \ccite{Coimbra:2013qq,Costa:2015llh}. \Scanners instead uses the
        analytic conditions (see \cref{app:pertun})
        \begin{equation}
          \begin{aligned}
            \frac{|\lambda|}{2}\eqcomma \;
            \frac{|\delta_2|}{2}\eqcomma \;
            \frac{|d_2|}{2}                                                                        & < 8\pi\eqcomma \\
            \frac{\left|2 d_2 + 3 \lambda \pm \sqrt{8 \delta_2^2 + {(2d_2-3\lambda)}^2}\right|}{4} & < 8\pi\eqdot
          \end{aligned}\label{eq:cxsmuni}
        \end{equation}
  \item Boundedness from below is ensured using the known analytic conditions~\cite{Coimbra:2013qq}.
  \item The oblique parameters are calculated and tested using the generic method described in \cref{sec:EWprecision}.
  \item The branching ratios of the scalars are calculated using
        \textsf{sHDECAY}~\cite{Costa:2015llh} through the \anyHdecay interface (see
        \cref{app:anyhdecay}).
  \item Predictions for gluon-fusion and $bb$-associated Higgs production at
        hadron colliders are obtained using tabulated results from
        \textsf{SusHi} (see \cref{sec:higgs}) and the parametrization included
        in \Hb is used to obtain cross sections for the $VH$-associated
        (sub)channels (see \ccite{Bechtle:2020pkv}).
  \item Constraints from Higgs searches and Higgs measurements are tested with
        \Hb and \Hs as described in \cref{sec:higgs} using the aforementioned
        predictions for branching ratios and cross sections.
  \item Using \Bsmpt for the calculation, the \ac{EWPT} can be required to be
        first order. Since this requirement is not a necessary constraint it is
        not enabled by default (see \cref{sec:EWPT}).
\end{itemize}

\subsubsection{The broken-phase \acs{CxSM} --- \texttt{CxSMBroken}}
The \ac{CxSM} is in the broken phase if both the real and imaginary parts of
$\mathbb{S}$ acquire non-zero \acp{vev}. This leads to a Higgs sector with three
mixing CP-even scalar bosons $h_i$ ($i\in\lbrace1,2,3\rbrace$). \Scanners uses the input
parameters~\cite{Costa:2015llh}
\begin{align}
    & m_a\eqcomma                &
    & m_b\eqcomma                &
    & \alpha^\text{in}_1\eqcomma &
    & \alpha^\text{in}_2\eqcomma &
    & \alpha^\text{in}_3\eqcomma &
    & v_S\eqcomma                &
  v & = v_\text{EW}\eqcomma
\end{align}
where $m_{a,b}$ are the masses of two of the $h_i$, $\alpha^\text{in}_{1,2,3}$
are the input mixing angles, $v_S$ is the real singlet \ac{vev}, and
$v_\text{EW}\approx\SI{246}{\GeV}$ is the \ac{EW} \ac{vev}. The third scalar
input mass $m_c$ is calculated through
\begin{equation}
  m_c^2 = - \frac{m_a^2m_b^2 R^\text{in}_{c1}R^\text{in}_{c2}}{m_a^2 R^\text{in}_{b1}R^\text{in}_{b2} + m_b^2 R^\text{in}_{a1}R^\text{in}_{a2}}\eqdot\label{eq:cxsmmasscalc}
\end{equation}
If \cref{eq:cxsmmasscalc} predicts a tachyonic $m_c^2$ the corresponding
parameter point is rejected.

\subsubsection{The dark-phase \acs{CxSM} --- \texttt{CxSMDark}}
If the imaginary part of $\mathbb{S}$ acquires no \ac{vev} during \ac{EWSB}, the
corresponding particle is a dark-matter candidate stabilized by a \Ztwo
symmetry. The real component of the singlet field still has a non-zero \ac{vev}
and mixes with the \ac{SM} Higgs boson. The additional assumption $a_1=0$ is
made, such that this dark phase of the \ac{CxSM} only has six real parameters:
\begin{align}
    & m_a\eqcomma           &
    & m_b\eqcomma           &
    & m_X\eqcomma           &
    & \alpha^\text{in}      &
    & v_S\eqcomma           &
  v & = v_\text{EW}\eqcomma
\end{align}
where $m_{a,b}$ are the masses of the two visible Higgs bosons, $m_X$ is the
mass of the dark scalar, $\alpha^\text{in}$ is the input mixing angle, and $v_S$
is the real singlet \ac{vev}. The convention for the mixing angle $\alpha$
matches \cref{eq:Rmix} with $\alpha_1\to\alpha$ and $\alpha_{2,3}\to0$ (and
analogously for the $\alpha^\text{in}$). Since the dark phase contains a \ac{DM}
candidate, the corresponding \ac{DM} constraints have to be considered in
addition to the constraints discussed in \cref{sec:CxSM}:
\begin{itemize}
  \item Dark matter observables are calculated using \Mo and tested against the
        experimental limits as discussed in \cref{sec:DM}.
\end{itemize}

\subsection{The Two-Real-Singlet-Extension of the SM}
The \ac{TRSM}~\cite{Robens:2019kga} is a different extension of the \ac{SM} by
two real scalar degrees of freedom. The scalar potential in terms of two real
scalar fields $S$ and $X$ is given by
\begin{equation}
  V_\text{TRSM} = \begin{aligned}[t]
     & \mu_{\Phi}^2 \Phi^\dagger \Phi + \lambda_{\Phi} {(\Phi^\dagger\Phi)}^2
    + \mu_{S}^2 S^2 + \lambda_S S^4
    + \mu_{X}^2 X^2 + \lambda_X X^4                                           \\
     & + \lambda_{\Phi S} \Phi^\dagger \Phi S^2
    + \lambda_{\Phi X} \Phi^\dagger \Phi X^2
    + \lambda_{XS} S^2 X^2\eqdot
  \end{aligned}
\end{equation}
This scalar potential respects a $\Ztwo\otimes\Ztwo$ symmetry for $S$ and $X$
and has nine real parameters. Depending on the vacuum structure the \ac{TRSM}
allows for different phases, though only the \emph{broken phase} is currently
implemented.

\subsubsection{The broken-phase \acs{TRSM} --- \texttt{TRSMBroken}}
In the broken phase both $S$ and $X$ acquire vacuum expectation values $v_S$ and
$v_X$, respectively. This phase with three mixing CP-even scalars $h_i$
($i\in\lbrace1,2,3\rbrace$) was studied in detail in \ccite{Robens:2019kga} and
we follow the conventions used there. The input parameters are
\begin{align}
   & M_a\eqcomma                                &
   & M_b\eqcomma                                &
   & M_c\eqcomma                                &
   & \theta^\text{in}_{hS}                      &
   & \theta^\text{in}_{hX}\eqcomma              &
   & \theta^\text{in}_{SX}\eqcomma              &
   & v_S\eqcomma                                &
   & v_X\eqcomma                                &
   & v = v_\text{EW}\eqcomma\label{eq:TRSMpars}
\end{align}
where $M_{a,b,c}$ are the three Higgs masses, $\theta^\text{in}_{hS}$,
$\theta^\text{in}_{hX}$, $\theta^\text{in}_{SX}$ are the input mixing angles,
and $v_{S,X}$ are the singlet \acp{vev}. The mixing angles $\theta$
parametrize the mixing matrix $R$ of \cref{eq:Rmix} as
\begin{align}
  \theta_{hS} & \equiv -\alpha_1\eqcomma &
  \theta_{hX} & \equiv -\alpha_2\eqcomma &
  \theta_{SX} & \equiv -\alpha_3\eqdot
\end{align}
Analogous relations hold for the corresponding input quantities
$\theta^\text{in}$, $R^\text{in}$, and $\alpha^\text{in}$.

In the broken phase of the \ac{TRSM}, \Scanners implements the following
constraints and calculations:
\begin{itemize}
  \item Perturbative unitarity and boundedness from below are ensured using the analytic
        conditions given in \ccite{Robens:2019kga}.
  \item The oblique parameters are calculated and tested using the generic
        method of \cref{sec:EWprecision}.
  \item The Higgs-to-Higgs decay widths of the scalars are calculated at
        tree-level and combined with the appropriately rescaled SM-like branching
        ratios as tabulated in \Hb using the \emph{effective coupling input} (see
        \ccite{Bechtle:2020pkv}).
  \item \Hb and \Hs are used to test constraints from Higgs data as described in
        \cref{sec:higgs}.
\end{itemize}

\subsection{The Two-Higgs-Doublet Model}\label{sec:2HDM}
The \ac{2HDM}~\cite{Lee:1973iz} (see \eg \ccite{Branco:2011iw} for a review) is
probably the most studied non-supersymmetric scalar extension of the \ac{SM}. In
the conventions of \ccite{Branco:2011iw} the scalar potential of the \ac{2HDM}
with a softly broken \Ztwo symmetry is given by
\begin{equation}
  V_\text{2HDM} =\begin{aligned}[t]
     & m_{11}^2 \Phi_1^\dagger\Phi_1
    + m_{22}^2 \Phi_2^\dagger\Phi_2
    - (m_{12}^2 \Phi_1^\dagger\Phi_2 + \hc)
    + \frac{\lambda_1}{2} {(\Phi_1^\dagger\Phi_1)}^2
    + \frac{\lambda_2}{2} {(\Phi_2^\dagger\Phi_2)}^2            \\
     & + \lambda_3 (\Phi_1^\dagger\Phi_1)(\Phi_2^\dagger\Phi_2)
    + \lambda_4 (\Phi_1^\dagger\Phi_2)(\Phi_2^\dagger\Phi_1)
    + (\frac{\lambda_5}{2}{(\Phi_1^\dagger\Phi_2)}^2 + \hc)
  \end{aligned}\label{eq:2hdmpot}
\end{equation}
The softly broken \Ztwo symmetry is extended to the Yukawa sector to prevent
tree-level flavour changing neutral currents leading to the four Yukawa types of
the \ac{2HDM}.

The parameters $m_{12}^2$ and $\lambda_5$ of \cref{eq:2hdmpot} can take complex
values and lead to a CP-violating scalar sector. \Scanners implements two
variants of the \ac{2HDM} --- the CP-conserving \ac{R2HDM} and the CP-violating
\ac{C2HDM}~\cite{Branco:1985aq,Ginzburg:2002wt,Khater:2003wq}. The following
constraints have similar implementations in both cases.
\begin{itemize}
  \item Perturbative unitarity and boundedness from below are tested using the
        analytic conditions given in \ccite{Branco:2011iw}.
  \item The oblique parameters are calculated and tested using the generic
        method of \cref{sec:EWprecision}.
  \item Constraints from $b$-physics are tested as discussed in
        \cref{sec:flavor}.
  \item Using \Bsmpt for the calculation, the \ac{EWPT} can be required to be
        first order. Since this requirement is not a necessary constraint it is
        not enabled by default (see \cref{sec:EWPT}).
\end{itemize}

\subsubsection{The \acs{R2HDM} --- \texttt{R2HDM}}
In the \ac{R2HDM}, CP-conservation is imposed in the Higgs sector forcing all
eight parameters of \cref{eq:2hdmpot} to be real. In the \ac{R2HDM}, the mixing
matrix is conventionally defined through
\begin{equation}
  \begin{pmatrix}
    H \\h
  \end{pmatrix} = \begin{pmatrix}
    \cos\alpha  & \sin\alpha \\
    -\sin\alpha & \cos\alpha \\
  \end{pmatrix}\begin{pmatrix}\rho_1\\\rho_2\end{pmatrix}\quad\text{with }-\pi/2\leq\alpha<\pi/2\eqcomma\label{eq:R2HDMmix}
\end{equation}
where $\rho_{1,2}$ are the real, neutral component fields of the two doublets.
One set of input parameters implemented in \Scanners is thus
\begin{align}
   & m_{H_a}\eqcomma          &
   & m_{H_b}\eqcomma          &
   & m_A\eqcomma              &
   & m_{H^\pm}\eqcomma        &
   & \alpha^\text{in}\eqcomma &
   & \tan\beta\eqcomma        &
   & m_{12}^2\eqcomma         &
   & v = v_\text{EW}\eqcomma
\end{align}
where $m_{H_{a,b}}$ are the neutral, CP-even Higgs masses, $m_A$ is the mass of
the pseudoscalar, $m_H^\pm$ is the charged Higgs mass, $\alpha^\text{in}$ is the
CP-even neutral sector input mixing angle, $\tan\beta=v_2/v_1$ is the ratio of
the \acp{vev}, and $m_{12}^2$ is the soft \Ztwo-breaking parameter. Furthermore,
the Yukawa type has to be specified, which can be either
\begin{align}
  1 & \equiv \text{type I}\eqcomma                  &
  2 & \equiv \text{type II}\eqcomma                 &
  3 & \equiv \text{lepton specific}\eqcomma         &
  4 & \equiv \text{flipped}\eqdot\label{eq:Yuktype}
\end{align}

Instead of using $\alpha^\text{in}$ as input parameter, \Scanners also implements an
alternative (default) input parametrization in terms of the effective gauge coupling of
$H_b$,
\begin{equation}
  c(H_b VV) = \sin(\beta - \alpha^\text{in})\eqdot
\end{equation}
Restricting $c(H_b VV)\sim 0$ forces $H_a$ to be close to the alignment limit.
This is the default parametrization of the \ac{R2HDM} consisting of
\begin{align}
   & m_{H_a}\eqcomma         &
   & m_{H_b}\eqcomma         &
   & m_A\eqcomma             &
   & m_{H^\pm}\eqcomma       &
   & c(H_b VV)\eqcomma       &
   & \tan\beta\eqcomma       &
   & m_{12}^2\eqcomma        &
   & v = v_\text{EW}\eqcomma
\end{align}
together with the Yukawa type.

The following constraints and calculations in the \ac{R2HDM} are implemented in
addition to the ones mentioned in \cref{sec:2HDM}:
\begin{itemize}
  \item Absolute stability of the \ac{EW} vacuum is ensured using the
        discriminant of \ccite{Barroso:2013awa}.
  \item The branching ratios of the scalars are calculated using
        \textsf{HDECAY}~\cite{Djouadi:1997yw,Harlander:2013qxa,Djouadi:2018xqq} through the
        \anyHdecay interface (see \cref{app:anyhdecay}).
  \item Predictions for gluon-fusion and $bb$-associated Higgs production at
        hadron colliders are obtained using tabulated results from
        \textsf{SUSHI} (see \cref{sec:higgs}).  The \Hb parametrizations are
        used to obtain cross section predictions for the $VH$-associated
        (sub)channels and for charged Higgs production in association with a
        top-quark (see \ccite{Bechtle:2020pkv}).
  \item \Hb and \Hs are used to test constraints from Higgs data as described in
        \cref{sec:higgs}.
\end{itemize}

\subsubsection{The \acs{C2HDM} --- \texttt{C2HDM}}
If complex values for $m_{12}^2$ and $\lambda_5$ are allowed, the \ac{2HDM}
scalar potential can be CP-violating~\cite{Ginzburg:2002wt}. The phases of these
two parameters are not both independent, such that the \ac{C2HDM} has nine real
free parameters. The \ac{C2HDM} was discussed in detail in
\ccite{Fontes:2017zfn} and we follow the conventions used there. The $3\times3$
mixing matrix of the neutral scalars is parametrized as in \cref{eq:Rmix}. The
input parametrization in terms of the mixing angles $\alpha^\text{in}_{1,2,3}$
is
\begin{align}
   & m_{H_a}\eqcomma            &
   & m_{H_b}\eqcomma            &
   & m_{H^\pm}\eqcomma          &
   & \alpha^\text{in}_1\eqcomma &
   & \alpha^\text{in}_2\eqcomma &
   & \alpha^\text{in}_3\eqcomma &
   & \tan\beta\eqcomma          &
   & \Re(m_{12}^2)\eqcomma      &
   & v = v_\text{EW}\eqcomma
\end{align}
where $m_{H_{a,b}}$ are two of the neutral Higgs masses, $m_{H^\pm}$ is the
charged Higgs mass, $\alpha^\text{in}_{1,2,3}$ are the neutral-sector input
mixing angles, $\tan\beta=v_2/v_1$ is the ratio of the \acp{vev} and
$\Re(m_{12}^2)$ is the real part of the soft \Ztwo-breaking parameter.
Furthermore, the Yukawa type needs to be specified. The third neutral scalar
mass $m_{H_c}$ is calculated from the two other ones using~\cite{Fontes:2017zfn}
\begin{equation}
  m_{H_c}^2 = \frac{m_{H_a}^2 R^\text{in}_{a3}(R^\text{in}_{a2}\tan\beta - R^\text{in}_{a1}) + m_{H_b}^2 R^\text{in}_{b3}(R^\text{in}_{b2}\tan\beta-R^\text{in}_{b1})}{R^\text{in}_{c3}(R^\text{in}_{c1}-R^\text{in}_{c2}\tan\beta)}\eqdot
\end{equation}
If this results in a tachyonic $H_c$ the parameter point is rejected.

\Scanners implements a second \ac{C2HDM} input parametrization in terms of the
couplings and mixing matrix elements of $H_a$ and $H_b$. The input parameters are
\begin{equation}
  \begin{aligned}[t]
     & m_{H_a}\eqcomma                         &
     & m_{H_b}\eqcomma                         &
     & m_{H^\pm}\eqcomma                       &
     & \tan\beta\eqcomma                       &
     & \Re(m_{12}^2)\eqcomma                   &
     & v = v_\text{EW}\eqcomma                   \\
     & c^2(H_a VV)\eqcomma                     &
     & |c(H_a t\bar{t})|^2\eqcomma             &
     & \mathrm{sign}(R^\text{in}_{a3})\eqcomma &
     & R^\text{in}_{b3}\eqcomma
  \end{aligned}
\end{equation}
where $c(H_a VV)$ denotes the effective coupling of $H_a$ to the massive gauge
bosons ($V\in\lbrace W^\pm,Z\rbrace$), $|c(H_a t\bar{t})|^2$ is the squared absolute value of
the effective coupling between $H_a$ and top-quarks, and $R^\text{in}_{a3}$ and
$R^\text{in}_{b3}$ are the mixing matrix elements between the input states
$H_{a,b}$ and the pseudoscalar gauge eigenstate. Additionally, the Yukawa type
has to be specified. The second sign required to compensate for the two squared
couplings as input is fixed by assuming
\begin{equation}
  c(H_a VV)c^e(H_a t\bar{t})>0\eqdot
\end{equation}
This assumption is enforced by the Higgs measurements for $H_a\equiv h_{125}$
which should be chosen when using this parametrization. The mixing angles are
obtained from these couplings using the relations
\begin{align}
  c(H_a VV)           & = \cos\beta R^\text{in}_{a1} + \sin\beta R^\text{in}_{a2}\eqcomma                               \\
  |c(H_a t\bar{t})|^2 & = {(c^e(H_a t\bar{t}))}^2 + {(c^o(H_a t\bar{t}))}^2                                             \\
                      & = {\left(\frac{R_{a2}}{\sin\beta}\right)}^2 + {\left(\frac{R_{a3}}{\tan\beta}\right)}^2\eqcomma
\end{align}
in combination with \cref{eq:Rmix}, where $c^e$ and $c^o$ refer to the CP-even
and CP-odd components of the effective Higgs-fermion coupling, respectively. If
this system of conditions does not yield a valid solution for
$\alpha^\text{in}_{1,2,3}$ the parameter point is rejected. This parametrization
is very useful to ensure that $H_a\equiv h_{125}$ has SM-like couplings and is
used by default.

\Scanners checks the following constraints in addition to the ones mentioned in
\cref{sec:2HDM}:
\begin{itemize}
  \item Absolute stability of the \ac{EW} vacuum is ensured using the
        discriminant of \ccite{Ivanov:2015nea}.
  \item The branching ratios of the scalars are calculated using
        \textsf{C2HDM\_HDECAY}~\cite{Fontes:2017zfn} through the \anyHdecay
        interface (see \cref{app:anyhdecay}).
  \item Predictions for gluon-fusion and $bb$-associated Higgs production at
        hadron colliders are obtained using tabulated results from
        \textsf{SUSHI} (see \cref{sec:higgs}). CP-mixing effects are included,
        but CP-interference effects like the ones discussed
        in~\cite{Liebler:2016ceh} are not included. The \Hb parametrizations are
        used to obtain cross section predictions for the CP-mixed
        $VH$-associated (sub)channels and for charged Higgs production in
        association with a top-quark (see \ccite{Bechtle:2020pkv}).
  \item \Hb and \Hs are used to test constraints from Higgs data as described in
        \cref{sec:higgs}.
  \item The model prediction for the electron \ac{EDM} is calculated
        following~\ccite{Abe:2013qla} and checked against the limit by the ACME
        collaboration~\cite{Andreev:2018ayy}.
\end{itemize}

\subsection{The Next-to \acs{2HDM}}\label{sec:N2HDM}
The \acf{N2HDM}~\cite{Chen:2013jvg,Drozd:2014yla,Muhlleitner:2016mzt} adds an
additional real scalar singlet field $\Phi_S$ to the \ac{R2HDM}. In the conventions of
\ccite{Muhlleitner:2016mzt} its scalar potential is given by
\begin{equation}
  V_\text{N2HDM}=\begin{aligned}[t]
     & m_{11}^2 \Phi_1^\dagger\Phi_1
    + m_{22}^2 \Phi_2^\dagger\Phi_2
    - m_{12}^2 (\Phi_1^\dagger\Phi_2 + \hc)
    + \frac{\lambda_1}{2} {(\Phi_1^\dagger\Phi_1)}^2
    + \frac{\lambda_2}{2} {(\Phi_2^\dagger\Phi_2)}^2            \\
     & + \lambda_3 (\Phi_1^\dagger\Phi_1)(\Phi_2^\dagger\Phi_2)
    + \lambda_4 (\Phi_1^\dagger\Phi_2)(\Phi_2^\dagger\Phi_1)
    + \frac{\lambda_5}{2}({(\Phi_1^\dagger\Phi_2)}^2 + \hc)     \\
     & + \frac{m_S^2}{2}\Phi_S^2
    + \frac{\lambda_6}{8}\Phi_S^4
    + \frac{\lambda_7}{2}(\Phi_1^\dagger\Phi_1)\Phi_S^2
    + \frac{\lambda_8}{2}(\Phi_2^\dagger\Phi_2)\Phi_S^2\eqdot
  \end{aligned}
\end{equation}
The \ac{N2HDM} scalar potential has an exact \Ztwo symmetry for the singlet
field in addition to the softly broken \Ztwo symmetry of the doublet field
inherited from the \ac{2HDM}. The doublet \Ztwo symmetry leads to the same
\ac{2HDM} types when extended to the Yukawa sector. The \ac{N2HDM} features
different phases, depending on which symmetries remain unbroken after \ac{EWSB}.
Most constraints and calculations are implemented similarly for all phases of
the \ac{N2HDM}:
\begin{itemize}
  \item Perturbative unitarity and boundedness from below~\cite{Klimenko:1984qx}
        are ensured using the analytic conditions given in
        \ccite{Muhlleitner:2016mzt}.
  \item The oblique parameters are calculated and tested using the generic
        method of \cref{sec:EWprecision}.
  \item The branching ratios of the scalars are calculated using
        \textsf{N2HDECAY}~\cite{Muhlleitner:2016mzt,Engeln:2018mbg} through the
        \anyHdecay interface (see \cref{app:anyhdecay}).
  \item Predictions for gluon-fusion and $bb$-associated Higgs production at
        hadron colliders are obtained using tabulated results from
        \textsf{SUSHI} (see \cref{sec:higgs}). The \Hb parametrizations are used
        to obtain cross section predictions for the $VH$-associated
        (sub)channels and for charged Higgs production in association with a
        top-quark (see \ccite{Bechtle:2020pkv}).
  \item \Hb and \Hs are used to test constraints from Higgs data as described in
        \cref{sec:higgs}.
  \item Metastability constraints on the stability of the \ac{EW} vacuum are
        obtained using the \Evade library. See also \ccite{Ferreira:2019iqb} for a
        detailed study of vacuum stability in the broken-phase \ac{N2HDM}.
\end{itemize}

\subsubsection{The broken-phase \acs{N2HDM} --- \texttt{N2HDMBroken}}
If both doublets and the singlet fields acquire non-zero \acp{vev} the \ac{EW}
vacuum of the \ac{N2HDM} is in the broken phase. In this phase, the three
CP-even neutral scalar fields mix with the $3\times3$ mixing matrix
parametrized as \cref{eq:Rmix}. The input parametrization in terms of the mixing
angles $\alpha^\text{in}_{1,2,3}$ is
\begin{equation}
  \begin{aligned}[t]
     & m_{H_a}\eqcomma            &
     & m_{H_b}\eqcomma            &
     & m_{H_c}\eqcomma            &
     & m_A\eqcomma                &
     & m_{H^\pm}\eqcomma          &
     & \tan\beta\eqcomma            \\
     & \alpha^\text{in}_1\eqcomma &
     & \alpha^\text{in}_2\eqcomma &
     & \alpha^\text{in}_3\eqcomma &
     & m_{12}^2\eqcomma           &
     & v_S\eqcomma                &
     & v = v_\text{EW}\eqcomma
  \end{aligned}
\end{equation}
where $m_{H_{a,b,c}}$ are the CP-even scalar Higgs masses, $m_A$ is the
pseudoscalar mass, $m_{H^\pm}$ is the charged Higgs mass, $\tan\beta=v_2/v_1$ is
the ratio of the doublet \acp{vev}, $\alpha^\text{in}_{1,2,3}$ are the input
mixing angles of the CP-even neutral scalar sector, $m_{12}^2$ is the soft
\Ztwo-breaking parameter and $v_S$ is the singlet \ac{vev}. Additionaly, the
Yukawa type has to be specified as in \cref{eq:Yuktype}. Similar to the
\ac{C2HDM} we provide a reparametrization in terms of effective couplings and
mixing matrix elements. In this case the input parameters are
\begin{equation}
  \begin{aligned}
     & m_{H_a}\eqcomma                       &
     & m_{H_b}\eqcomma                       &
     & m_{H_c}\eqcomma                       &
     & m_A\eqcomma                           &
     & m_{H^\pm}\eqcomma                     &
     & \tan\beta\eqcomma                       \\
     & c^2(H_a VV)\eqcomma                   &
     & c^2(H_a t\bar{t})\eqcomma             &
     & \text{sign}(R^\text{in}_{a3})\eqcomma &
     & R^\text{in}_{b3}                      &
     & m_{12}^2\eqcomma                      &
     & v_S\eqcomma                           &
     & v = v_\text{EW}\eqcomma
  \end{aligned}
\end{equation}
where $c(H_a VV)$ and $c(H_a t\bar{t})$ are the effective couplings of $H_a$ to
massive gauge bosons and top-quarks, respectively, while $R^\text{in}_{a3}$ and
$R^\text{in}_{b3}$ are the mixing matrix elements between $H_{a,b}$ and the
singlet field. The Yukawa type has to be specified as an additional input
parameter. The second sign required to compensate for the squared coupling input
is again fixed by the assumption
\begin{equation}
  c(H_a VV)c(H_a t\bar{t})>0\eqcomma
\end{equation}
which is physically motivated for $H_a\equiv h_{125}$ which should be chosen
when using this parametrization. We obtain the mixing angles from the couplings
using
\begin{align}
  c(H_a VV)       & = \cos\beta R^\text{in}_{a1} + \sin\beta R^\text{in}_{a2}\eqcomma \\
  c(H_a t\bar{t}) & = \frac{R^\text{in}_{a2}}{\sin\beta}\eqcomma
\end{align}
and \cref{eq:Rmix}.

On top of the constraints listed in \cref{sec:N2HDM}, the broken phase
\ac{N2HDM} implements the following:
\begin{itemize}
  \item Using \Bsmpt for the calculation~\cite{Basler:2019iuu}, the \ac{EWPT}
        can be required to be first order. Since this requirement is not a
        necessary constraint it is not enabled by default (see \cref{sec:EWPT}).
\end{itemize}

\subsubsection{The dark-singlet-phase \acs{N2HDM} --- \texttt{N2HDMDarkS}} If
the singlet \ac{vev} is zero after \ac{EWSB}, the singlet field $\Phi_S$ is a
mass eigenstate stabilized by the \Ztwo symmetry of $\Phi_S$. In this
dark-singlet-phase~\cite{Drozd:2014yla, Engeln:2018ywp, Engeln:2018mbg,
  Engeln:2020fld} we follow the conventions of \ccite{Engeln:2020fld}. The mixing
between the two visible CP-even neutral scalars is treated in the convention of
the \ac{R2HDM} in \cref{eq:R2HDMmix}, though --- for better analogy to the other
phases --- this $2\times2$ mixing is embedded in a $3\times3$ mixing
matrix~\cite{Engeln:2020fld}. The input parameters are
\begin{equation}
  \begin{aligned}[t]
     & m_{H_a}\eqcomma          &
     & m_{H_b}\eqcomma          &
     & m_A\eqcomma              &
     & m_{H^\pm}\eqcomma        &
     & m_{H_D}\eqcomma          &
     & \tan\beta\eqcomma          \\
     & \alpha^\text{in}\eqcomma &
     & m_{12}^2\eqcomma         &
     & \lambda_6\eqcomma        &
     & \lambda_7\eqcomma        &
     & \lambda_8\eqcomma        &
     & v = v_\text{EW}\eqcomma
  \end{aligned}
\end{equation}
together with the Yukawa type, where now $m_{H_{a,b}}$ are the masses of the
visible CP-even scalars,  $m_{H_D}$ is the mass of the dark scalar, and
$\alpha^\text{in}$ is the remaining visible-sector mixing angle. In addition to
the constraints listed in \cref{sec:N2HDM}, the dark-singlet-phase \ac{N2HDM}
also implements the following:
\begin{itemize}
  \item Dark matter observables are calculated using \Mo and tested against the
        experimental limits as discussed in \cref{sec:DM}. Due to the
        limitations of the model file format used by \Mo, dark matter
        observables can currently only be calculated for Yukawa sectors of
        type~I.
\end{itemize}

\subsubsection{The dark-doublet-phase \acs{N2HDM} --- \texttt{N2HDMDarkD}}
If $m_{12}^2=0$ it is possible that only one Higgs doublet of the \ac{N2HDM}
acquires a \ac{vev}. The resulting exact \Ztwo symmetry leads to an inert Higgs
doublet similar to the inert doublet model~\cite{Deshpande:1977rw}. As long as
the singlet \ac{vev} is non-zero, there is still mixing between the \ac{SM}-like
doublet Higgs and the singlet. We again follow the conventions of
\ccite{Engeln:2020fld}. The \Scanners input parameters in the dark-doublet-phase
\ac{N2HDM} are
\begin{align}
   & m_{H_a}\eqcomma          &
   & m_{H_b}\eqcomma          &
   & m_{A_D}\eqcomma          &
   & m_{H_D^\pm}\eqcomma      &
   & m_{H_D}\eqcomma          &
   & \alpha^\text{in}\eqcomma &
   & m_{22}^2\eqcomma         &
   & \lambda_2\eqcomma        &
   & \lambda_8\eqcomma        &
   & v_S\eqcomma              &
   & v = v_\text{EW}\eqcomma
\end{align}
where $m_{A_D}$ and $m_{H^D}$ denote the masses of the two opposite-CP neutral
dark scalars and $m_{H_D^\pm}$ is the dark charged scalar mass. Since one of the
doublets is inert, there are no Yukawa types in this phase. The
dark-doublet-phase \ac{N2HDM} implements the following additional constraints:
\begin{itemize}
  \item Dark matter observables are calculated using \Mo and tested against the
        experimental limits as discussed in \cref{sec:DM}.
\end{itemize}

\subsubsection{The fully-dark-phase \acs{N2HDM} --- \texttt{N2HDMDarkSD}}
If the singlet \ac{vev} additionally vanishes, both \Ztwo symmetries of the
\ac{N2HDM} remain exact after \ac{EWSB}. This phase features a
visible sector identical to the \ac{SM} and two distinct dark sectors --- one
containing the inert doublet and the other composed of just the singlet --- that
are also stabilized with respect to each other. Since there is no mixing in this
phase, the input parametrization is extremely simple. In the conventions of
\ccite{Engeln:2020fld}, \Scanners uses the parameters
\begin{align}
   & m_{H_\text{SM}}\eqcomma &
   & m_{H^D_D}\eqcomma       &
   & m_{A_D}\eqcomma         &
   & m_{H_D^\pm}\eqcomma     &
   & m_{H^S_D}\eqcomma       &
   & m_{22}^2\eqcomma        &
   & m_S^2                   &
   & \lambda_2\eqcomma       &
   & \lambda_6\eqcomma       &
   & \lambda_8\eqcomma       &
   & v = v_\text{EW}\eqcomma
\end{align}
where $H^D_D$ is the real, neutral dark Higgs from the inert doublet sector, and
$H^S_D$ is the dark singlet. The fully-dark \ac{N2HDM} also implements \ac{DM}
constraints:
\begin{itemize}
  \item Dark matter observables are calculated using \Mo and tested against the
        experimental limits as discussed in \cref{sec:DM}. The \Mo routines
        adapted for two-component \ac{DM} are used.
\end{itemize}

\subsection{Minimal CP-Violating Dark Matter --- \texttt{CPVDM}}
The minimal model of CP-violating scalar dark matter~\cite{Azevedo:2018fmj} is a
variant of the \ac{N2HDM}, where the two separate \Ztwo symmetries for the
doublet and singlet fields are merged into one. The resulting scalar potential
in the conventions of \ccite{Azevedo:2018fmj} is
\begin{equation}
  V_\text{CPVDM}=\begin{aligned}[t]
     & m_{11}^2 \Phi_1^\dagger\Phi_1
    + m_{22}^2 \Phi_2^\dagger\Phi_2
    + \frac{m_S^2}{2}\Phi_S^2
    + \left(A \Phi_1^\dagger \Phi_2\Phi_S + \hc\right)                 \\
     & + \frac{\lambda_1}{2} {(\Phi_1^\dagger\Phi_1)}^2
    + \frac{\lambda_2}{2} {(\Phi_2^\dagger\Phi_2)}^2                   \\
     & + \lambda_3 (\Phi_1^\dagger\Phi_1)(\Phi_2^\dagger\Phi_2)
    + \lambda_4 (\Phi_1^\dagger\Phi_2)(\Phi_2^\dagger\Phi_1)
    + \frac{\lambda_5}{2}\left({(\Phi_1^\dagger\Phi_2)}^2 + \hc\right) \\
     & + \frac{\lambda_6}{8}\Phi_S^4
    + \frac{\lambda_7}{2}(\Phi_1^\dagger\Phi_1)\Phi_S^2
    + \frac{\lambda_8}{2}(\Phi_2^\dagger\Phi_2)\Phi_S^2\eqcomma
  \end{aligned}
\end{equation}
where the trilinear parameter $A$ can be complex.\footnote{A priori, $\lambda_5$
  could also be complex, but it can always be made real by a rephasing of the
  fields.} Only $\Phi_1$ acquires a \ac{vev} such that the visible sector is
identical to the \ac{SM}, but CP-violating mixing between the three neutral
fields in the dark sector is induced by the $A$ term. With the mixing matrix
parametrized as in \cref{eq:Rmix} the \Scanners input parameters are
\begin{equation}
  \begin{aligned}
     & m_{h}\eqcomma                                               &
     & m_{h_a}\eqcomma                                             &
     & m_{h_b}\eqcomma                                             &
     & m_{H^\pm}\eqcomma                                           &
     & \alpha^\text{in}_1\eqcomma                                  &
     & \alpha^\text{in}_2\eqcomma \quad \alpha^\text{in}_3\eqcomma   \\
     & \lambda_2\eqcomma                                           &
     & \lambda_6\eqcomma                                           &
     & \lambda_8\eqcomma                                           &
     & m_{22}^2\eqcomma                                            &
     & m_S^2\eqcomma                                               &
     & v=v_\text{EW}\eqcomma
  \end{aligned}
\end{equation}
where $h$ is the \ac{SM}-like visible Higgs boson and $h_{a,b}$ are dark neutral
scalars. The third dark neutral Higgs mass $m_{h_c}$ is calculated through the
relation
\begin{equation}
  m_{h_c}^2 = -\frac{m_{h_a}^2 R_{a1} R_{a2} + m_{h_b}^2 R_{b1}  R_{b2}}{R_{c1} R_{c2}}\eqdot
\end{equation}
The parameter point is rejected if this leads to a tachyonic $h_c$. The
CP violation in this model does not enter the fermion sector and, therefore, no
CP-sensitive \ac{EDM} constraints need to be considered for the model. The
implemented constrains are:
\begin{itemize}
  \item Perturbative unitarity and boundedness from below are ensured using the same
        analytic conditions as in the \ac{N2HDM}~\cite{Muhlleitner:2016mzt}.
  \item The oblique parameters are calculated and tested using the generic
        method of \cref{sec:EWprecision}.
  \item The Higgs-to-Higgs decay widths of the scalars are calculated at
        tree-level and combined with the appropriately rescaled SM-like branching
        ratios as tabulated in \Hb using the \emph{effective coupling input} (see
        \ccite{Bechtle:2020pkv}).
  \item \Hb and \Hs are used to test constraints from Higgs data as described in
        \cref{sec:higgs}.
  \item Dark matter observables are calculated using \Mo and tested against the
        experimental limits as discussed in \cref{sec:DM}.
  \item Metastability constraints on the stability of the \ac{EW} vacuum are
        obtained using the \Evade library. See also \ccite{Ferreira:2019iqb} for a
        detailed study of vacuum stability in the broken-phase \ac{N2HDM}.
\end{itemize}

\section{User Operating Instructions}\label{sec:user}
\lstset{language=bash,basicstyle=\ttfamily}
The \Scanners source code is available at
\begin{center}
  \url{https://gitlab.com/jonaswittbrodt/ScannerS}\eqdot{}
\end{center}
\Scanners requires the following tools and libraries to be available on the
system:
\begin{itemize}
  \item While most of the code is written in \texttt{C++}, some of the
        dependencies are \texttt{Fortran} codes such that working compilers for
        \texttt{C}, \texttt{C++}, and \texttt{Fortran} are required. The
        \texttt{C++} compiler has to support at least \texttt{C++-17}.
  \item \Scanners is compiled using \textsf{CMake} which has to be available on
        the system.
  \item The \textsf{GSL}~\cite{Galassi2009} is used for some numerical operations.
  \item The \textsf{Eigen3} library~\cite{Guennebaud2010} is used throughout the code.
\end{itemize}
Detailed information on the required versions is included in the \texttt{README}
file.

If these basic requirements are fulfilled, \Scanners can be compiled using
\begin{lstlisting}
  mkdir build && cd build
  cmake ..
  make
\end{lstlisting}
after which the \Scanners test suite can be run with \texttt{make test}. The
physics codes required by \Scanners are automatically downloaded by
\textsf{CMake}. These currently include \Hbcite, \Hscite, and \anyHdecay (see
\cref{app:anyhdecay}). Additionally, there are several optional dependencies to
enable additional constraints.
\begin{itemize}
  \item \Mocite is needed for the calculation of \ac{DM} observables.
  \item \Evade~\cite{Hollik:2018wrr, Ferreira:2019iqb, EVADEweb} is needed for
        metastability constraints.
  \item \Bsmpt~\cite{Basler:2018cwe, Basler:2020nrq} is needed to compute the
        order and strength of the \ac{EW} phase transition.
\end{itemize}
Again, see the \texttt{README} file for more technical information on how to link
these codes to \Scanners.

Compiling \Scanners generates one executable in the build directory for each
implemented model. The executable names for each model are given in the title of
the corresponding subsection in \cref{sec:models}. All of the \Scanners
executables support two run modes: \emph{scan} mode and \emph{check} mode. In
\emph{scan} mode, \Scanners generates and tests random parameter points until
the requested number of valid points is found. It can be called as \eg
\begin{lstlisting}
  ./R2HDM --config example_input/R2HDM_T1.ini scan -n 10
\end{lstlisting}
to generate ten valid parameter points using the scan ranges in the example
configuration file in the \ac{R2HDM} and write them into an output file with the
default name \texttt{R2HDM.tsv} (the same name as the executable with the
extension \texttt{.tsv} added). All possible options are documented in the
command line help which can be accessed \eg through \texttt{./R2HDM --help}
(these include constraint severities and input parameter ranges). The
configuration file simply specifies command line options. Since \Scanners
performs a random scan, it is easily parallelizable by launching multiple jobs
with the same input and different output files.

The output format used by \Scanners is a simple tab-separated tabular format
where each line corresponds to one parameter point. The first column contains an
index which identifies the point, while all following columns contain data. The
first line contains a header that specifies the contents of each
row.\footnote{The row names are documented with the function that stores the
  corresponding value to the output. This can either be one of the constraints in
  \href{https://jonaswittbrodt.gitlab.io/ScannerS/namespaceScannerS_1_1Constraints.html}{\texttt{ScannerS::Constraints}}
  or a member function of the model class in
  \href{https://jonaswittbrodt.gitlab.io/ScannerS/namespaceScannerS_1_1Models.html}{\texttt{ScannerS::Models}}.}
The \textsf{pandas} Python package for example can be used to easily read:
\lstset{language=Python,basicstyle=\ttfamily}
\begin{lstlisting}
  df = pandas.read_table("R2HDM.tsv", index_col=0)
\end{lstlisting}
and write:
\begin{lstlisting}
  df.to_csv("R2HDM_copy.tsv", sep="\t")
\end{lstlisting}
files in this format.

In \emph{check} mode, the input parameters are read from a file. All of the
input parameter points that fulfill the constraints are written to the output.
This can \eg be used to re-check a set of parameter points when the constraints
have been updated. For example
\lstset{language=bash,basicstyle=\ttfamily}
\begin{lstlisting}
  ./R2HDM R2HDM_re.tsv check R2HDM.tsv
\end{lstlisting}
would re-check the parameter points generated in the previous example. Of
course, unless the constraints changed between the two runs, \texttt{R2HDM.tsv}
and \texttt{R2HDM\_re.tsv} would contain the same parameter points. The
\emph{check} mode can also be used to apply the constraints to a manually
specified set of parameter points, \eg a parameter plane of interest. In models
with multiple implemented sets of input parameters, \emph{check} mode always
uses the simplest possible input method --- usually the input in terms of mixing
angles --- to minimize the numerical errors induced by repeated re-checks.

\subsection{Extending \Scanners}
\Scanners can easily be extended given some \texttt{C++} programming experience.
New constraints can be added by extending the minimal example given in the
documentation of the
\href{https://jonaswittbrodt.gitlab.io/ScannerS/classScannerS_1_1Constraints_1_1Constraint.html}{\texttt{ScannerS::Constraints::Constraint}}
class. The existing constraints illustrate a wide variety of possibilities. Some
have almost trivial implementations that delegate all calculations to the model
--- such as the boundedness (\texttt{ScannerS::Constraints::BFB}) and unitarity
(\texttt{ScannerS::Constraints::Unitarity}) constraints --- some perform
calculations themselves --- \eg the constraint
\texttt{ScannerS::Constraints::STU} that calculates and checks the oblique
parameters --- or delegate to external codes --- like
\texttt{ScannerS::Constraints::Higgs}, the constraint from Higgs observables,
that calls \Hb and \Hs.

New models are also straightforward to implement. The main requirement is to
implement the relations between the chosen set of input parameters and the
remaining model parameters --- \eg the expressions for the $\lambda_i$ in terms
of masses and mixing angles --- in the constructor of a \texttt{ParameterPoint}
member class. The basic technical requirements on a model class are listed in
the documentation of the
\href{https://jonaswittbrodt.gitlab.io/ScannerS/namespaceScannerS_1_1Models.html}{\texttt{Scanners::Models}}
namespace.

Constraints are enabled for a model class by implementing their required
functions and attributes. In most cases these are static member functions of the
model class operating on a \texttt{ParameterPoint} object, the detailed
requirements can be found in the documentation of each constraint in the
\href{https://jonaswittbrodt.gitlab.io/ScannerS/namespaceScannerS_1_1Constraints.html}{\texttt{Scanners::Constraints}}
namespace.

\section{Summary}\label{sec:summary}
We have presented the \texttt{C++} code \Scanners that performs parameter scans
in many \ac{BSM} models with extended Higgs sectors. The resulting samples of
parameter points can be used \eg for phenomenological studies, as benchmark
scenarios for experimental searches, or as numerical examples for precision
calculations.

\Scanners implements many different sources of constraints on the models. These
include theoretical constraints, constraints from precision measurements in the
EW and flavour sector, constraints from LHC Higgs data, \ac{DM} constraints, and
\ac{EDM} constraints as well as the requirement of a strong-first-order EW phase
transition, if applicable. These constraints can be applied in parameter scans
of many different implemented \ac{BSM} models --- different singlet extensions
of the \ac{SM}, both the CP-conserving and CP-violating \ac{2HDM}, and many
different phases and variants of the \ac{N2HDM}.

\Scanners aims to be easy to install --- with automatic dependency management
for the physics codes it uses --- and easy to use --- using a straightforward
command line interface and simple tabular data files. We have used \Scanners to
generate tens of millions of valid parameter points across the implemented
models and the code proved to be reliable and efficient.

Implementing new constraints or models in \Scanners is straightforward and
encouraged. The online documentation specifies the technical requirements and
most extensions can be easily implemented by modifying and extending existing
model or constraint implementations. If you have implemented new models or
constraints in \Scanners we encourage you to contact us with regards to merging
them back into the main code.

\subsection*{Acknowledgements}
We thank Duarte Azevado, Philipp Basler, Thomas Biekötter, Isabell Engeln, and
Jonas Müller for using and testing pre-release versions of \Scanners and for
providing important feedback. MM is supported by the BMBF-Project 05H18VKCC1. RS
acknowledges FCT support, Contracts UIDB/00618/2020, UIDP/00618/2020,
PTDC/FIS-PAR/31. JW acknowledges funding by the European Research Council (ERC)
under the European Union's Horizon 2020 research and innovation program, grant
agreement No 668679.

\appendix
\section{The \anyHdecay Library}\label{app:anyhdecay}

In order to apply constraints from Higgs searches and Higgs measurements,
precise model predictions for Higgs branching ratios and total widths are
needed.\footnote{In simple models --- such as pure singlet extensions --- this
  can be circumvented by using the \Hb \emph{effective coupling
    input}~\cite{Bechtle:2020pkv}.} \Scanners uses the interface library \anyHdecay to
obtain these predictions from dedicated tools based on the code
\textsf{HDECAY}~\cite{Djouadi:1997yw, Harlander:2013qxa, Djouadi:2018xqq} that
exist for various models. As these codes are not designed to be used as a
library, \anyHdecay wraps their functionality into a common \texttt{C++}
interface.

The library currently incorporates \textsf{HDECAY} for the \ac{SM} and
\ac{R2HDM}, \textsf{sHDECAY}~\cite{Costa:2015llh} for the different phases of
the \ac{CxSM} (and the real-singlet-extended SM (RxSM) that is not currently
implemented in \Scanners),
\textsf{N2HDECAY}~\cite{Muhlleitner:2016mzt,Engeln:2018mbg} for the phases of
the \ac{N2HDM}, and \textsf{C2HDM\_HDECAY}~\cite{Fontes:2017zfn} for the
\ac{C2HDM}. The \anyHdecay source code is available at
\begin{center}
  \url{https://gitlab.com/jonaswittbrodt/anyhdecay}
\end{center}
and only requires working \texttt{Fortran} and \texttt{C++-17} compilers as well
as \textsf{CMake}. The underlying codes are downloaded through \textsf{CMake}
and automatically restructured to allow linking them into one library. The API
documentation for \anyHdecay is available at
\begin{center}
  \url{https://jonaswittbrodt.gitlab.io/anyhdecay/}\eqdot{}
\end{center}

\section{Perturbative Unitarity Bounds}\label{app:pertun}
\lstset{language=Mathematica,basicstyle=\ttfamily}
The tree-level $2\to2$ scattering matrix $\mathcal{M}_{2\to2}$ in the high
energy limit is easy to derive directly from the scalar potential. An element
of $\mathcal{M}_{2\to2}$ corresponding to a set of two particle states
$\ket{AB}$ and $\ket{CD}$ is given by
\begin{equation}
  \braket{AB | \mathcal{M} | CD} = \frac{1}{\sqrt{(1+\delta_{AB})(1+\delta_{CD})}} \diffp{V}{ABCD}\eqcomma \label{eq:scattermatrix}
\end{equation}
where the $\delta$ functions lead to the correct symmetry factors for the
two-particle states. The perturbative unitarity constraint bounds the $n$
eigenvalues $\mathcal{M}^i_{2\to2}$ ($i\in\lbrace1,\ldots,n\rbrace$) of
$\mathcal{M}_{2\to2}$ as
\begin{equation}
  |\mathcal{M}^i_{2\to2}|<8\pi\eqdot
\end{equation}
However, since basis transformations are unitary transformations, the
$\mathcal{M}^i_{2\to2}$ are independent of the basis~\cite{Kanemura:1993hm}, and
the most convenient one --- usually the basis of gauge eigenstates --- can be
used for the calculation.

The eigenvalues of $\mathcal{M}_{2\to2}$ can be calculated numerically in a
straightforward way. However, in most cases the diagonalization can --- at least
partially --- be carried out analytically. This both leads to additional insight
into the allowed ranges for the parameters of the scalar potential and allows a
much faster evaluation of perturbative unitarity constraints. A small
\Mathematica package --- \texttt{tools/PerturbativeUnitarity.m} --- that can be
used to perform this calculation is included in \Scanners. We will illustrate
its usage by deriving the \ac{CxSM} unitarity constraints of \cref{eq:cxsmuni}.

The package is loaded as usual (with the path adjusted as needed)
\begin{lstlisting}
  <<tools/PerturbativeUnitarity.m
\end{lstlisting}
It only needs the quartic part of the scalar potential --- in terms of
eigenstates of electric charge --- and a list of fields as input. For the
\ac{CxSM} we define the quartic part of the scalar potential, \cref{eq:cxsmpot},
as
\begin{lstlisting}
  Phi = {{1/Sqrt[2] (HCr    + I HCi)},
         {1/Sqrt[2] (v + HR + I HI )}};
  PhiCC = ConjugateTranspose[Phi];
  PhiS = (vsR + sR + I vsI + I sI)/Sqrt[2];

  hChargedTrans = {HCr ->   (Hm + Hp)/Sqrt[2],
                   HCi -> I (Hm - Hp)/Sqrt[2]};

  V = Expand[
        ComplexExpand[
          lambda/4 (PhiCC.Phi)^2
          + delta2/2 (PhiCC.Phi) Abs[PhiS]^2
          + d2/4 Abs[PhiS] ^4
        ][[1, 1]] /. hChargedTrans];
\end{lstlisting}
where the \texttt{hChargedTrans} rules are used to transform from the real field
components into the complex electrically charged fields. The fields are defined
as
\begin{lstlisting}
  fields = {{Hp, Hm}, HR, HI, sR, sI};
\end{lstlisting}
where any charged fields should be given in a two-component sub-list with the
positive eigenstate first and the negative eigenstate second. This information
is used to avoid generating redundant charge conjugate states. With this input
the package generates the scatter matrix $\mathcal{M}_{2\to2}$ through
\begin{lstlisting}
  (sMat = scatterMatrix[V, fields]) // scatterMatrixForm
\end{lstlisting}
The \texttt{scatterMatrixForm} creates a formatted output of the --- in case of
the \ac{CxSM} $16\times16$ --- scattering matrix, where the corresponding
two-particle states are indicated at each row and column. This full scatter
matrix is decomposed into sub-blocks using
\begin{lstlisting}
  blocks = splitBlockMatrix[sMat];
  scatterMatrixForm /@ blocks // TableForm
\end{lstlisting}
The second line prints all of the sub-matrices. In the \ac{CxSM} this will show
a $1\times1$ matrix of the doubly-charged states $\ket{H^\pm H^\pm}$, four
$1\times1$ matrices for the singly-charged states, and another six $1\times1$
matrices and one $5\times5$ matrix for the neutral states. The eigenvalues
$\mathcal{M}^i_{2\to2}$ are then obtained using
\begin{lstlisting}
  uniqueEV[blocks]
\end{lstlisting}
which gives the exact results of \cref{eq:cxsmuni}. The maximum and minimum
values for the quartic potential parameters that are possible without violating
unitarity may also be of interest. These can be obtained using
\begin{lstlisting}
  maxUnitaryParRange[uniqueEV[blocks], {lambda,delta2,d2}]
\end{lstlisting}
where the second argument lists the quartic parameters that enter the unitarity
constraints. For the \ac{CxSM} this returns that perturbative unitarity can only
be fulfilled if
\begin{align}
  |\lambda|  & \lesssim 16.8\eqcomma &
  |\delta_2| & \lesssim 35.6\eqcomma &
  |d_2|      & \lesssim 25.2\eqdot
\end{align}
This information is particularly relevant to obtain reasonable scan ranges for
models where some of the quartic parameters are input parameters --- such as the
dark phases of the \ac{N2HDM}. The function works by numerically minimizing and
maximizing each parameter given the perturbative unitarity constraints on all
parameters. Optionally, additional constraints that should be included can be
given as a third argument. For example, by including the \ac{CxSM} boundedness
conditions~\cite{Coimbra:2013qq} as
\begin{lstlisting}
  maxUnitaryParRange[
    uniqueEV[blocks],
    {lambda,delta2,d2},
    lambda > 0
      && d2 > 0
      && (delta2^2 < lambda d2 || delta2 > 0)]
\end{lstlisting}
we obtain the smaller maximally allowed parameter ranges
\begin{align}
   & 0      \leq     \lambda  \lesssim 16.8\eqcomma &
   & -13.1  \lesssim \delta_2 \lesssim 35.6\eqcomma &
   & 0      \leq     d_2      \lesssim 25.2\eqdot
\end{align}

For more complicated models such as the \ac{N2HDM} it is not possible to obtain
all eigenvalues analytically. In this case some of the $\mathcal{M}^i_{2\to2}$
in the output of the \texttt{uniqueEV} command will be expressed as the roots of
a polynomial. These roots can then be calculated numerically, which is still
substantially simpler than diagonalizing the full scatter matrix numerically ---
in the \ac{N2HDM} the difference is between finding the roots of a cubic
polynomial and diagonalizing a $40\times40$ scatter matrix. This exploding
complexity is the reason, why \Scannersv{2} no longer uses the generic numerical
method for perturbative unitarity constraints~\cite{Coimbra:2013qq} and instead
includes this \Mathematica package for easily obtaining (semi-)analytic
conditions.

\printbibliography{}

\end{document}